\batchmode

\documentstyle[12pt]{article}
\newcommand{\nc}{\newcommand}
\nc{\renc}{\renewcommand}

\nc{\half}{{\textstyle{1\over2}}}
\nc{\etal}{\mbox{\it et al. }}
\nc{\ie}{{\it i.e.}}
\nc{\eg}{{\it e.g.}}

\renc{\thefootnote}{\arabic{footnote}}
\nc{\capt}[1]{{\bf Figure.} {\small\sl #1}}

\nc{\eqs}[2]{\mbox{Eqs.~(\ref{#1},\,\ref{#2})}}
\nc{\eq}[1]{\mbox{Eq.~(\ref{#1})}}

\nc{\figs}[2]{\mbox{Figs.~(\ref{#1},\,\ref{#2})}}
\nc{\fig}[1]{\mbox{Fig~.(\ref{#1})}}

\nc{\tag}[1]{\label{#1} \marginpar{{\footnotesize #1}}}
\nc{\mtag}[1]{\label{#1} \mbox{\marginpar{{\footnotesize #1}}}}
\renc{\baselinestretch}{1.2}
\newlength{\overeqskip}
\newlength{\undereqskip}
\nc{\be}[1]{\begin{equation} \mbox{$\label{#1}$}}
\nc{\bea}[1]{\begin{eqnarray} \mbox{$\label{#1}$}}
\nc{\Section}[2]{\section{#2}\label{#1}}
\nc{\Bibitem}[1]{\bibitem{#1}}
\nc{\Label}[1]{\label{#1}}

\nc{\eea}{\vspace{\undereqskip}\end{eqnarray}}
\nc{\ee}{\vspace{\undereqskip}\end{equation}}
\nc{\bdm}{\begin{displaymath}}
\nc{\edm}{\end{displaymath}}
\nc{\dpsty}{\displaystyle}
\nc{\bc}{\begin{center}}
\nc{\ec}{\end{center}}
\nc{\ba}{\begin{array}}
\nc{\ea}{\end{array}}
\nc{\bab}{\begin{abstract}}
\nc{\eab}{\end{abstract}}
\nc{\btab}{\begin{tabular}}
\nc{\etab}{\end{tabular}}
\nc{\bit}{\begin{itemize}}
\nc{\eit}{\end{itemize}}
\nc{\ben}{\begin{enumerate}}
\nc{\een}{\end{enumerate}}
\nc{\bfig}{\begin{figure}}
\nc{\efig}{\end{figure}}
\nc{\arreq}{&\!=\!&}
\nc{\arrmi}{&\!-\!&}
\nc{\arrpl}{&\!+\!&}
\nc{\arrap}{&\!\!\!\approx\!\!\!&}
\nc{\non}{\nonumber\\*}
\nc{\align}{\!\!\!\!\!\!\!\!&&}

\def\lsim{\; \raise0.3ex\hbox{$<$\kern-0.75em
      \raise-1.1ex\hbox{$\sim$}}\; }
\def\gsim{\; \raise0.3ex\hbox{$>$\kern-0.75em
      \raise-1.1ex\hbox{$\sim$}}\; }
\nc{\DOT}{\hspace{-0.08in}{\bf .}\hspace{0.1in}}
\nc{\Laada}{\hbox {$\sqcap$ \kern -1em $\sqcup$}}
\nc\loota{{\scriptstyle\sqcap\kern-0.55em\hbox{$\scriptstyle\sqcup$}}}
\nc\Loota{{\sqcap\kern-0.65em\hbox{$\sqcup$}}}
\nc\laada{\Loota}
\nc{\qed}{\hskip 3em \hbox{\BOX} \vskip 2ex}

\nc{\real}{{\rm I \! R}}
\nc{\Z}{{\sf Z \!\!\! Z}}
\nc{\complex}{{\rm C\!\!\! {\sf I}\,\,}}
\def\bigid{\leavevmode\hbox{\small1\kern-3.8pt\normalsize1}}
\def\id{\leavevmode\hbox{\small1\kern-3.3pt\normalsize1}}
\nc{\slask}{\!\!\!/}
\nc{\bis}{{\prime\prime}}
\nc{\pa}{\partial}
\nc{\na}{\nabla}
\nc{\ra}{\rangle}
\nc{\la}{\langle}
\nc{\goto}{\rightarrow}
\nc{\swap}{\leftrightarrow}

\nc{\EE}[1]{ \mbox{$\cdot10^{#1}$} }
\nc{\abs}[1]{\left|#1\right|}
\nc{\at}[2]{\left.#1\right|_{#2}}
\nc{\norm}[1]{\|#1\|}
\nc{\abscut}[2]{\Abs{#1}_{\scriptscriptstyle#2}}
\nc{\vek}[1]{{\rm\bf #1}}
\nc{\integral}[2]{\int\limits_{#1}^{#2}}
\nc{\inv}[1]{\frac{1}{#1}}
\nc{\dd}[2]{{{\partial #1}\over{\partial #2}}}
\nc{\ddd}[2]{{{{\partial}^2 #1}\over{\partial {#2}^2}}}
\nc{\dddd}[3]{{{{\partial}^2 #1}\over
	{\partial #2 \partial #3}}}
\nc{\dder}[2]{{{d #1}\over{d #2}}}
\nc{\ddder}[2]{{{d^2 #1}\over{d {#2}^2}}}
\nc{\dddder}[3]{{d^2 #1}\over
	{d #2 d #3}}
\nc{\dx}[1]{d\,^{#1}x}
\nc{\dy}[1]{d\,^{#1}y}
\nc{\dz}[1]{d\,^{#1}z}
\nc{\dl}[1]{\frac{d\,^{#1}l}{(2\pi)^{#1}}}
\nc{\dk}[1]{\frac{d\,^{#1}k}{(2\pi)^{#1}}}
\nc{\dq}[1]{\frac{d\,^{#1}q}{(2\pi)^{#1}}}

\nc{\cc}{\mbox{$c.c.$ }}
\nc{\hc}{\mbox{$h.c.$ }}
\nc{\cf}{cf.\ }
\nc{\erfc}{{\rm erfc}}
\nc{\Tr}{{\rm Tr\,}}
\nc{\tr}{{\rm tr\,}}
\nc{\pol}{{\rm pol}}
\nc{\sign}{{\rm sign}}
\nc{\bfT}{{\bf T }}

\def\GeV{{\rm\ GeV}}
\def\MeV{{\rm\ MeV}}
\def\keV{{\rm\ keV}}
\def\TeV{{\rm\ TeV}}

\nc{\cA}{{\cal A}}
\nc{\cB}{{\cal B}}
\nc{\cD}{{\cal D}}
\nc{\cE}{{\cal E}}
\nc{\cG}{{\cal G}}
\nc{\cH}{{\cal H}}
\nc{\cL}{{\cal L}}
\nc{\cO}{{\cal O}}
\nc{\cT}{{\cal T}}
\nc{\cN}{{\cal N}}
\nc{\rvac}[1]{|{\cal O}#1\rangle}
\nc{\lvac}[1]{\langle{\cal O}#1|}
\nc{\rvacb}[1]{|{\cal O}_\beta #1\rangle}
\nc{\lvacb}[1]{\langle{\cal O}_\beta #1 |}
\nc{\bb}{\bar{\beta}}
\nc{\bt}{\tilde{\beta}}
\nc{\ctH}{\tilde{\cal H}}
\nc{\chH}{\hat{\cal H}}
\nc{\1}{\aa}
\nc{\2}{\"{a}}
\nc{\3}{\"{o}}
\nc{\4}{\AA}
\nc{\5}{\"{A}}
\nc{\6}{\"{O}}
\nc{\al}{\alpha}
\nc{\g}{\gamma}
\nc{\Del}{\Delta}
\nc{\e}{\epsilon}
\nc{\eps}{\epsilon}
\nc{\lam}{\lambda}
\nc{\om}{\omega}
\nc{\Om}{\Omega}
\nc{\ve}{\varepsilon}
\nc{\mn}{{\mu\nu}}
\nc{\k}{\kappa}
\nc{\vp}{\varphi}

\nc{\advp}[3]{{\it  Adv.\ in\ Phys.\ }{{\bf #1} {(#2)} {#3}}}
\nc{\annp}[3]{{\it  Ann.\ Phys.\ (N.Y.)\ }{{\bf #1} {(#2)} {#3}}}
\nc{\apl}[3]{{\it  Appl. Phys. Lett. }{{\bf #1} {(#2)} {#3}}}
\nc{\apj}[3]{{\it  Ap.\ J.\ }{{\bf #1} {(#2)} {#3}}}
\nc{\apjl}[3]{{\it  Ap.\ J.\ Lett.\ }{{\bf #1} {(#2)} {#3}}}
\nc{\app}[3]{{\it Astropart.\ Phys.\ }{{\bf #1} {(#2)} {#3}}}
\nc{\cmp}[3]{{\it  Comm.\ Math.\ Phys.\ }{{ \bf #1} {(#2)} {#3}}}
\nc{\cqg}[3]{{\it  Class.\ Quant.\ Grav.\ }{{\bf #1} {(#2)} {#3}}}
\nc{\epl}[3]{{\it  Europhys.\ Lett.\ }{{\bf #1} {(#2)} {#3}}}
\nc{\ijmp}[3]{{\it Int.\ J.\ Mod.\ Phys.\ }{{\bf #1} {(#2)} {#3}}}
\nc{\ijtp}[3]{{\it Int.\ J.\ Theor.\ Phys.\ }{{\bf #1} {(#2)} {#3}}}
\nc{\jmp}[3]{{\it  J.\ Math.\ Phys.\ }{{ \bf #1} {(#2)} {#3}}}
\nc{\jpa}[3]{{\it  J.\ Phys.\ A\ }{{\bf #1} {(#2)} {#3}}}
\nc{\jpc}[3]{{\it  J.\ Phys.\ C\ }{{\bf #1} {(#2)} {#3}}}
\nc{\jap}[3]{{\it J.\ Appl.\ Phys.\ }{{\bf #1} {(#2)} {#3}}}
\nc{\jpsj}[3]{{\it J.\ Phys.\ Soc.\ Japan\ }{{\bf #1} {(#2)} {#3}}}
\nc{\lmp}[3]{{\it Lett.\ Math.\ Phys.\ }{{\bf #1} {(#2)} {#3}}}
\nc{\mpl}[3]{{\it  Mod.\ Phys.\ Lett.\ }{{\bf #1} {(#2)} {#3}}}
\nc{\ncim}[3]{{\it  Nuov.\ Cim.\ }{{\bf #1} {(#2)} {#3}}}
\nc{\np}[3]{{\it  Nucl.\ Phys.\ }{{\bf #1} {(#2)} {#3}}}
\nc{\pr}[3]{{\it Phys.\ Rev.\ }{{\bf #1} {(#2)} {#3}}}
\nc{\pra}[3]{{\it  Phys.\ Rev.\ A\ }{{\bf #1} {(#2)} {#3}}}
\nc{\prb}[3]{{\it  Phys.\ Rev.\ B\ }{{{\bf #1} {(#2)} {#3}}}}
\nc{\prc}[3]{{\it  Phys.\ Rev.\ C\ }{{\bf #1} {(#2)} {#3}}}
\nc{\prd}[3]{{\it  Phys.\ Rev.\ D\ }{{\bf #1} {(#2)} {#3}}}
\nc{\prl}[3]{{\it Phys.\ Rev.\ Lett.\ }{{\bf #1} {(#2)} {#3}}}
\nc{\pl}[3]{{\it  Phys.\ Lett.\ }{{\bf #1} {(#2)} {#3}}}
\nc{\prep}[3]{{\it Phys.\ Rep.\ }{{\bf #1} {(#2)} {#3}}}
\nc{\prsl}[3]{{\it Proc.\ R.\ Soc.\ London\ }{{\bf #1} {(#2)} {#3}}}
\nc{\ptp}[3]{{\it  Prog.\ Theor.\ Phys.\ }{{\bf #1} {(#2)} {#3}}}
\nc{\ptps}[3]{{\it  Prog\ Theor.\ Phys.\ suppl.\ }{{\bf #1} {(#2)} {#3}}}
\nc{\physa}[3]{{\it  Physica\ A\ }{{\bf #1} {(#2)} {#3}}}
\nc{\physb}[3]{{\it  Physica\ B\ }{{\bf #1} {(#2)} {#3}}}
\nc{\phys}[3]{{\it Physica\ }{{\bf #1} {(#2)} {#3}}}
\nc{\rmp}[3]{{\it  Rev.\ Mod.\ Phys.\ }{{\bf #1} {(#2)} {#3}}}
\nc{\rpp}[3]{{\it Rep.\ Prog.\ Phys.\ }{{\bf #1} {(#2)} {#3}}}
\nc{\sjnp}[3]{{\it Sov.\ J.\ Nucl.\ Phys.\ }{{\bf #1} {(#2)} {#3}}}
\nc{\spjetp}[3]{{\it Sov.\ Phys.\ JETP\ }{{\bf #1} {(#2)} {#3}}}
\nc{\yf}[3]{{\it Yad.\ Fiz.\ }{{\bf #1} {(#2)} {#3}}}
\nc{\zetp}[3]{{\it Zh.\ Eksp.\ Teor.\ Fiz.\  }{{\bf #1}  {(#2)} {#3}}}
\nc{\zp}[3]{{\it Z.\ Phys.\ }{{\bf #1} {(#2)} {#3}}}
\nc{\ibid}[3]{{\sl ibid.\ }{{\bf #1} {#2} {#3}}}
\nc{\rf}[1]{(\ref{#1})}
\nc{\nn}{\nonumber \\*}
\nc{\bfB}{\bf{B}}
\nc{\bfv}{\bf{v}}
\nc{\bfx}{\bf{x}}
\nc{\bfy}{\bf{y}}
\nc{\vx}{\vec{x}}
\nc{\vy}{\vec{y}}
\nc{\oB}{\overline{B}}
\nc{\oI}{\overline{I}}
\nc{\oR}{\overline{R}}
\nc{\rar}{\rightarrow}
\nc{\ti}{\times}
\nc{\slsh}{\hskip-5pt/}
\nc{\sm}{Standard~Model~}
\nc{\MP}{M_{\rm Pl}}
\nc{\tp}{t_{\rm Pl}}
\nc{\ave}{\bar{E}}

\nc{\eff}{{\rm eff}}
\nc{\kk}{\vek{k}}
\nc{\pp}{{\rm p}}
\nc{\ga}{g_{a\gamma}}
\nc{\vv}{\\}
\nc{\eee}{{\bf E}}
\nc{\bbb}{{\bf B}}
\nc{\qcd}{T_{\rm QCD}}
\nc{\G}{\rm \ G}
\def\vec#1{{\bf #1}}

\def\lae{\;^{<}_{\sim} \;} \def\gae{\; ^{>}_{\sim} \;} 

\def\udd{u^{c}d^{c}d^{c}}

\def\uude{u^{c}u^{c}d^{c}e^{c}}

\begin{document}
{\title{\vskip-2truecm{\hfill {{\small \\
	}}\vskip 1truecm}
{\bf Q-Balls and Baryogenesis in the MSSM}}
{\author{
{\sc  Kari Enqvist$^{1}$ and John McDonald$^{2}$}\\
{\sl\small Department of Physics, P.O. Box 9,
FIN-00014 University of Helsinki,
Finland}
}
\maketitle
\vspace{1cm}
\begin{abstract}
\noindent
	We show that Q-balls naturally exist in the
 Minimal Supersymmetric Standard Model (MSSM) 
with soft SUSY breaking terms of the minimal N=1 SUGRA type.
 These are associated with the F- and D-flat directions 
of the scalar potential once radiative corrections are taken into
 account. We consider two
 distinct cases, corresponding to the
 "$H_{u}L$" (slepton) direction with L-balls and the "$u^{c}d^{c}d^{c}$" and "$\uude$"
(squark) directions with B-balls.
 The L-ball always has a small charge, typically of the order of 1000, whilst
 the B-ball can have an arbitrarily large charge, which,
when created cosmologically by the collapse of an unstable
 Affleck-Dine condensate, is likely to be greater than
$10^{14}$. The B-balls typically decay at
 temperatures less than that of the electroweak phase transition, leading
 to a novel version of Affleck-Dine baryogenesis, 
in which the B asymmetry comes from
 Q-ball decay rather than condensate decay. This mechanism 
can work even in the presence of additional L violating 
interactions or $B-L$ conservation, which would 
rule out conventional Affleck-Dine baryogenesis. 
\end{abstract}
\vfil
\footnoterule
{\small $^1$enqvist@pcu.helsinki.fi};
{\small $^2$mcdonald@outo.helsinki.fi}

\thispagestyle{empty}
\newpage
\setcounter{page}{1}

\section{Introduction}

                        Flat directions in the scalar potential of the Minimal Supersymmetric
 Standard Model (MSSM) \cite{nilles} have long been understood to have potentially important
 consequences for cosmology \cite{ad,drt,fd}. In 
particular, a very natural mechanism for the generation
 of the observed baryon asymmetry in SUSY models is the 
decay of a charged scalar field
 condensate (the Affleck-Dine (AD) mechanism \cite{ad}). 
 Developments in the cosmology of N=1 supergravity models, in particular 
the generation of order $H$ contributions to the soft SUSY breaking parameters
 of scalar fields in the early Universe \cite{hc}, have
 sparked a revival of interest in flat directions 
and the AD mechanism \cite{drt}. This mechanism
 is particularly important as the other natural possibility 
for baryogenesis in the MSSM, namely electroweak
 baryogenesis, appears to succeed for only a very
limited range of MSSM parameters \cite{ewb}.
 A second idea has recently been reconsidered in the context of SUSY models
 with flat directions in their scalar potentials, namely Q-balls [7-11].
 A Q-ball is a stable, charge Q soliton in a scalar field theory with a spontaneously
 broken global $U(1)$ symmetry \cite{cole}. The
 connection between Q-balls and SUSY
 models lies in the natural existence of 
scalars carrying a global $U(1)$ charge, corresponding to $B$ or $L$ for the case
 of squarks and sleptons. Thus, in principle, one could have B- or L-balls in 
the MSSM. Whether this happens or not depends
 on whether the scalar potential of the squarks
 and sleptons can support a Q-ball solution. If
 it can, then the next questions are whether
 Q-balls can be created in the early Universe and whether
 the resulting Q-balls could be sufficiently long-lived to have interesting consequences
 for cosmology. 
In reference \cite{ks} it was shown, in the context of a
 SUSY model with gauge mediated SUSY breaking, 
that the scalar potential along a flat direction of the MSSM
 can support a Q-ball solution. In addition,
 it was suggested that Q-balls with a very large charge (and so possibly stable) could be
 formed by the growth of perturbations in a charged condensate
 of the kind associated with the AD mechanism.

                In this letter we will reconsider these ideas in the
 context of the "classical" MSSM, i.e. the MSSM with 
soft SUSY
 breaking terms due to SUSY breaking in a gravitationally coupled hidden sector \cite{nilles}. 
We will first show that the scalar potential along the
 flat "$H_{u}L$" (slepton) and "$u^{c}d^{c}d^{c}$" 
and "$\uude$"
 (squark) directions can indeed support a Q-ball solution
 once radiative corrections are taken into account.
 The form of the potential and the associated Q-balls
 will be seen to be quite different in the slepton and 
squark cases. 
We will then discuss the formation of Q-balls in the early Universe in
 the context of the simplest cosmological scenario, 
in which the AD field begins coherently oscillating 
during the inflaton matter dominated era following an initial period 
of inflation. Finally, we will discuss the lifetime of the Q-balls and 
their possible consequences for cosmology, in particular 
their role in baryogenesis.

\section{Q-balls and radiatively corrected flat directions in the MSSM}
\subsection{Flat directions in the MSSM}

               For a globally $U(1)$ symmetric complex scalar field $\phi$, 
the condition for the scalar potential $U(\phi)$ to
 be able to support a Q-ball solution is that $U(\phi)/|\phi|^2$
 should have a global minimum 
at a non-zero value of $|\phi|$ \cite{cole}. 
 In this it is assumed that $\phi$ is
 defined such that $\phi = 0$ corresponds to the vacuum in which the Q-ball
 is formed, which in practice
corresponds to our vacuum with non-zero Higgs expectation values and zero
 squark and slepton expectation values. We will be interested
 in whether Q-balls can be formed along
flat directions of the MSSM. 
The flat directions can be defined by the scalar field operators which have non-zero 
expectation values along the directions in question.
 These have been classified generally for the case
 of the MSSM in reference \cite{drt}. The most
 interesting flat directions from the point of view of
 conventional AD baryogenesis are those with non-zero B-L and 
unbroken R-parity, which can
 give a baryon asymmetry once
 anomalous
 B+L violation is taken into account without introducing dangerous 
renormalizable
 B and L violating operators in the MSSM \cite{nilles,zwirner}. 
These flat directions correspond to the "$H_{u}L$" direction, along which 
the magnitude of the 
expectation value of $H_{u}^{0}$ and 
$\nu_{L}$ are equal, and the "$u^{c}d^{c}d^{c}$" direction, along
 which $u_{c}$, $d_{c}$ and $d_{c}^{'}$ are non-zero, where the
 squarks have different colour indices and where $d_{c}$ and
 $d_{c}^{'}$ correspond to orthogonal combinations of down squark
 generations. (There are also the "$d^{c}QL$" and "$e^{c}LL$" directions. 
However, since we
expect these to be phenomenologically similar to
 the "$\udd$" direction, we will not consider
 them seperately here). R-parity conservation allows 
the d=4 non-renormalizible superpotential term $(H_{u}L)^{2}$ and the d=6 term 
$(\udd)^{2}$. In addition, we will consider the d=4
 B-L conserving $\uude$ direction (there is also the 
phenomenologically similar $QQQL$ direction). Although
 this could not produce a B asymmetry via
AD condensate decay once anomalous $B+L$
 violation is taken into account, we will see
 that it can generate a baryon asymmetry via Q-ball decays
 occuring after the electroweak phase transition. 

              For sufficiently large $|\phi|$ along the D-flat direction the scalar potential is
of the form
\be{pot1} U(\phi) = m_{S}^{2} |\phi|^{2} 
+ \frac{\lambda^{2} |\phi|^{2(d-1)}}{M_{p}^{2(d-3)}} 
+ ( \frac{A_{\lambda} \lambda \phi^{d}}{M_{p}^{d-3}} + h.c.)
 ~,\ee
where $m_{S}^{2}$ is the soft SUSY breaking mass squared term and 
$A_{\lambda}$ the A-term, and 
we take the non-renormalizible terms to be suppressed by the Planck scale $M_{p}$. 
Thus, if $m_{S}$ has no
 $|\phi|$ dependence, then $U(\phi)/|\phi|^{2}$ will not have a global minimum at
 non-zero $|\phi|$ and so no stable Q-ball will exist. In order to have
 such a $|\phi|$ dependence, Kusenko and Shaposhnikov \cite{ks}
considered the possibility that $m_{S}^{2}$ is generated by gauge
 mediated SUSY breaking. In this case, for $|\phi|$ larger than the mass
 of the messenger quarks, $m_{S}$ 
becomes proportional to $|\phi|^{-2}$ and the potential becomes 
essentially completely flat up to where the non-renormalizible terms become important. 
However, gauge mediated SUSY breaking is not the most commonly
 considered form of SUSY breaking. The most common
 SUSY breaking mechanism is that where
soft SUSY breaking terms are generated by SUSY breaking in
 a hidden sector, not coupled directly to the observable MSSM fields in the
 K\"ahler potential of N=1 supergravity \cite{nilles}. This results in soft SUSY breaking
 terms which are constant at tree level. However, once radiative corrections
 are taken into account, they become scale dependent. Indeed, it is widely believed
 that radiative corrections to the soft SUSY breaking mass squared term of the Higgs
 scalar giving masses to the up-type quarks, $H_{u}$, when summed from
 the GUT or Planck scale using the renormalization
 group (RG), are responsible for driving its mass squared negative at renormalization scales
 of the order of the weak scale, so breaking the electroweak symmetry \cite{nilles,rewb}. It is
 well-known that the effect of radiative corrections on the 
effective potential as a function of $|\phi|$ can be summed by replacing
 the tree-level masses and couplings by their value run by the RG
 equations to a renormalization scale $\mu 
\approx |\phi|$ \cite{effpot}. Thus, for sufficiently small values of $|\phi|$, the scalar potential 
 along the D-flat direction will have the form $U(\phi) \approx m_{S}^{2}(|\phi|)|\phi|^{2}$.
A necessary condition for 
this potential to be able to support a Q-ball solution 
is that $m_{S}^{2}(|\phi|)$ decreases over at least some range of $|\phi|$ as $|\phi|$
 increases from zero. 

\subsection{$H_{u}L$ direction}

        In general we will be considering models with several scalar fields. 
To correctly describe the Q-balls, which correspond to L-balls 
in this case, we would have to solve the coupled
equations of motion of the scalars to find a Q-ball solution. However, in practice, we will
be able to describe the nature of the Q-balls reasonably accurately by considering a single
scalar degree of freedom. This can be understood as follows. 
Along the D-flat direction, for large values of the scalar field
compared with the mass scale of the soft SUSY breaking terms
(which are typically of the order of $100\GeV$), the degree of freedom in the direction
orthogonal to the D-flat direction will be much more
 massive than $100\GeV$ and so will play no role in the 
dynamics of the Q-ball. Thus we need only consider the
 scalar field with non-zero value
along the D-flat direction. However, at sufficiently small values of the scalar field,
 the orthogonal scalar
 mass becomes of the same order
as the soft SUSY breaking masses and so all the scalar
 fields can play a role in the dynamics of the Q-ball.
However, in this limit, all the fields will typically have similar masses
 and couplings and so we can simply consider
 one of the scalars to be decoupled from all the others and
 calculate the Q-ball dynamics for this scalar. 
This will give the correct order of magnitude for the properties of the Q-ball 
in this regime. 

           Let us first consider the potential at large enough
 values of $|\phi|$ that the $H_{u}L$ D-flat direction describes the minimum
 of the potential. In this case the scalar field along the minimum of the potential 
corresponds to a linear
 combination of the $H_{u}^{o}$ and $\nu_{L}$ fields,
 $\phi = \frac{1}{\sqrt{2}}(H_{u}^{o} + \nu_{L})$,
 with mass $m_{S}^{2} = \frac{1}{2}(\mu_{H}^{2} + m_{H_{u}}^2
+ m_{L}^{2}) $, where $\mu_{H}$ corresponds
 to the SUSY $\mu$-term in the superpotential, 
$\mu_{H} H_{u}H_{d}$. In general, the renormalization group equations for the 
soft SUSY breaking mass terms will 
have the form
\be{nilles} \mu \frac{\partial m_{i}}{\partial \mu} = \alpha_{i}
 m_{i}^{2} -\beta_{\alpha} M_{\alpha}^{2}  ~,\ee
where the $m_{i}^{2}$ represent the soft SUSY breaking
 scalar mass terms as well as the A-terms, and
$M_{\alpha}$ are the gaugino masses. The full RG 
equations are given in reference \cite{nilles}. The only important 
$\alpha_{i}$ terms are those associated with the top quark
 Yukawa coupling. These appear in the RG equation 
for $m_{H_{u}}$ but not for $m_{L}$. In the absence of a large 
$\alpha_{i}$ the masses increase monotonically with decreasing $\mu$. 
However, with a large $\alpha_{i}$, for sufficiently
 small $\mu$, the masses can start to
decrease with decreasing $\mu$. 

               The form of the solutions depends on the initial values of the
 parameters of the SUSY breaking terms at the initial large scale, which 
we will take in the following to be the 
GUT scale $M_X$. 
However, two features are important for the existence and cosmology 
of L-balls.
 First, in general, the value of $m_{S}^{2}$ becomes negative 
for scales typically smaller than $10^{8}$ GeV or so. Second, depending
 on the parameters, there can be a "hill" in 
the plot of $m_{S}^{2}$ versus $|\phi|$, such that 
$m_{S}^{2}$ starts decreasing with increasing
 $|\phi|$ for sufficiently large values of $|\phi|$. 

The running of the mass in the $H_uL$-direction as
obtained from solving the coupled one-loop RG equations
is illustrated in Fig. 1a for  different initial values of the SUSY 
parameters, chosen so that they give radiative
 breaking of electroweak symmetry
at the scale $\mu\simeq M_W$ (we took $\alpha_s(M_W)=0.11$ and the top
Yukawa $h_t(M_W)=1.05$; no threshold corrections were augmented).
We have not made a systematic exploration of the parameter space, but
it appears that typically the ``hill'' in the plot of $m_{S}^{2}$ 
versus $|\phi|$ arises for $m_0$ smaller than the gaugino masses or
for $A(M_X)<0$. Examples of both can be seen in Fig. 1a.

\begin{figure}
\leavevmode
\centering
\vspace*{70mm} 
\includegraphics{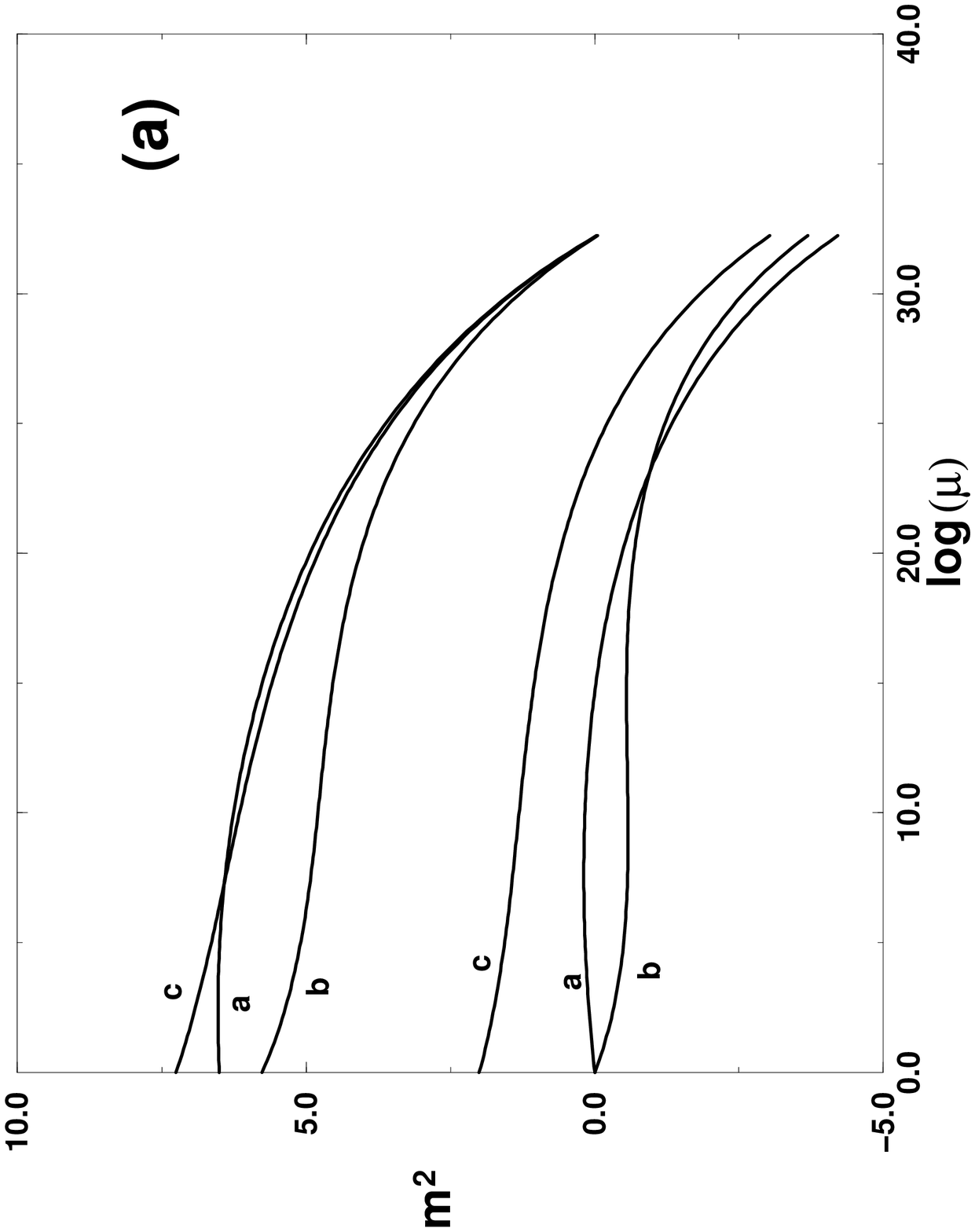}
\includegraphics{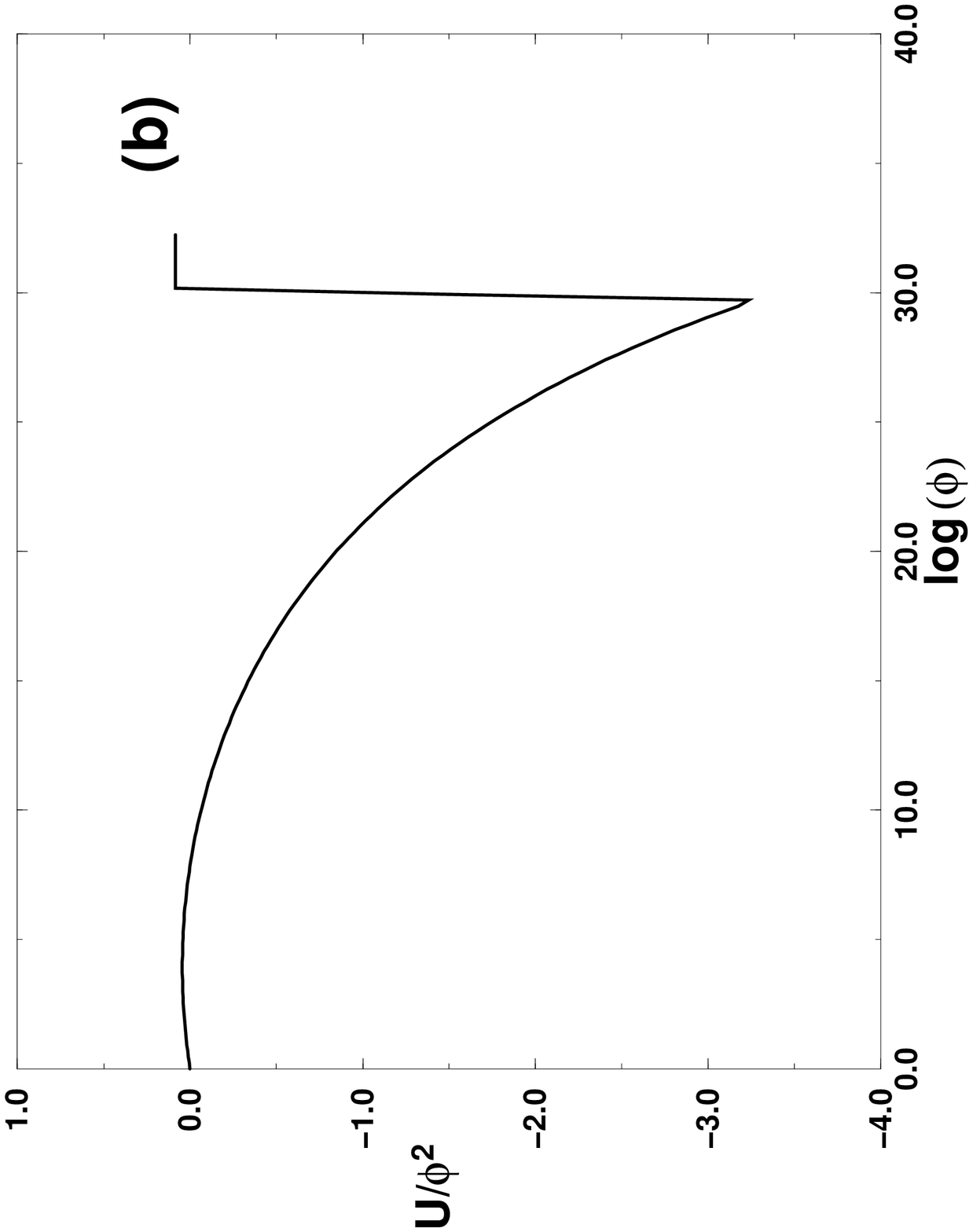}   
\caption{{\bf a)} The running of the mass in the $H_uL$ direction (lower
set of curves) for the susy parameters (a) $m_0=0,~A(M_X)=0,~\mu_H=2.55$,
(b) $m_0=0,~A(M_X)=-3,~\mu_H=2.4$
 and (c) $m_0=1,~A(M_X)=-1,~\mu_H=2.5$,
all in units of the common gaugino mass; for comparision,
the running of the physical Higgs mass is also shown (upper set of curves).
{\bf b)} The Q-ball potential for the case (a).}
\label{kuva1}       
\end{figure} 

             The effect of the negative value of $m_{S}^{2}$ at small enough 
$|\phi|$ is to generate a minimum of $U(|\phi|)/|\phi|^{2}$, typically
 at $|\phi_{o}| \approx 1\TeV$. 
To see this, we first note that the definition of the L-ball field $\phi$ 
will not correspond to the D-flat direction once $H_{u}^{o}
 \lae m_{L}/g$. The $\nu_{L}$ expectation value vanishes at the minimum 
of the potential as a function of $H_{u}^{o}$ in this case. In this regime the
 L-ball field  may be taken to be 
$\nu_{L}$. This will have a positive mass squared,
 corresponding to our L-conserving vacuum. 
Thus we see that the effective L-ball field will
 be characterized by a positive mass squared for 
$|\phi| \lae 1\TeV$ and a negative mass squared for $|\phi| \gae 1\TeV$.
The rate by which the positive
mass switches on is given by the rate by which the Higgs VEV develops.
This mass turns positive over a very narrow range in $\phi$.
This is illustrated in Fig. 1b.  We need not be concerned with the details 
of the potential; the L-ball properties will be essentially determined by
this rapid change of the sign of the effective mass squared of the L-ball 
field, which is responsible for 
$U(|\phi|)/|\phi|^{2}$ having a minimum at $|\phi_{o}| \approx 1\TeV$. 
 
         The potential importance of the "hill" feature in Fig. 1b  
is that if a condensate
 in the early Universe starts to oscillate coherently with its initial
 amplitude in the range of $|\phi|$ where $m_{S}^{2}$ is decreasing with
 increasing $|\phi|$, then, as we will show later, there could, in principle, 
be a negative pressure
associated with the condensate, which would lead to the exponential
growth of perturbations in the condensate field and 
ultimately to the formation of L-balls. 

\subsection{$u^{c}d^{c}d^{c}$ direction}

             In this case the D-flat direction essentially describes the B-ball field $\phi$
 for all values of $|\phi|$. The corresponding mass term will have the form $m_{S}^{2}
 \approx \alpha_{i} m_{u \; i}^{2} + \beta_{i} m_{d\;i}^{2} + 
\gamma_{i} m_{d^{'}\;i}^{2}$, where the sum is over different generations $i$ and 
where $d$ and $d^{'}$ correspond to different colours and orthogonal
 combinations of down squark generations, 
in order to have a non-zero value for the operator 
$\epsilon^{ijk}u^{c}_{i}d^{c}_{j}d^{c\;'}_{k}$, where i, j and k are colour indices. 
The values of $\alpha_{i}$, 
$\beta_{i}$ and $\gamma_{i}$ will be determined
 by the initial conditions following inflation,
 which will 
determine the direction in colour and flavour space of the flat
 direction. On average, in the sense of 
allowing equal contributions from all squark species, we
 expect that the effective mass will be of the form 
$m_{S}^{2} \approx  \frac{1}{9}(m_{t^{c}}^{2} +
 2 m_{u^{c}}^{2}  + 6 m_{d^{c}}^{2}) $ (where we assume a common mass for 
first and second generation up squarks and 
for all down squarks), although
 it is quite possible to have, for example, a much smaller contribution from 
$m_{t^{c}}^{2}$ (which can $decrease$
 with decreasing $|\phi|$ for small enough $|\phi|$)
 for particular initial conditions. We generally expect $m_{S}^{2}$ to decrease
 with increasing $|\phi|$ without any hill feature or change of sign,
in complete contrast with the case of the $H_{u}L$ direction. 
This leads to a correction to the 
$\phi$ mass which we can roughly model as a logarithmic correction
\be{logc} m_{S}^{2} \approx m_{o}^{2} \left(1 + K \log\left(\frac{|\phi|^{2}}
{M^{2}_X}\right) \right)~,
\ee
where $K < 0$ and $M_X$ is a large mass scale at which
 $m_{S}^{2}$ is defined to have the value $m_{o}^{2}$. 
Perturbatively, the value of $K$ due to SUSY breaking gaugino masses will be of the form
\be{pertk}   K \approx - \sum_{\alpha,\; gauginos}
 \frac{\alpha_{g_{\alpha}}}{8 \pi} \frac{M_{\alpha}^{2}}{m_{S}^{2}}  ~\ee
which, taking $\alpha_{g_{\alpha}} \approx 0.1$ for $SU(3)_{c}$ and summing over the 
gauginos gaining a $\phi$ dependent mass from 
mixing with the $u^{c}$ and $d^{c}$ squarks,
 will typically give $|K|$ in the range 0.01 to 0.1.    
The minimum of the
 potential will then be determined by the non-renormalizible terms in the potential. 
The form of the non-renormalizible terms will depend on 
the flat direction in question.
 Since we are considering
R-parity to be unbroken, the lowest-order non-renormalizible
superpotential term lifting the scalar potential along the $\udd$ direction 
will be of the form $\frac{\lambda}{M_{p}^{3}} (\udd)^2$. Thus
 the full scalar potential in this direction will have the form
\be{uddsp} U(\phi) \approx m_{o}^{2} \left(1 + K \log \left(\frac{|\phi|^{2}}
{M^{2}_X}\right) \right) |\phi|^{2} 
+ \frac{\lambda^{2}}{M_{p}^{6}} |\phi|^{10} ~.
\ee
In what follows we will always take $\lambda\approx 1$. Thus
the value of $|\phi|$ at the minimum of $U(|\phi|)/|\phi|^{2}$ is given by 
$|\phi|_{o} \approx (|K| m_{S}^{2} M_{Pl}^{6})^{1/8} \approx 5 \times
10^{14} |K|^{1/8}\GeV$ for 
$m_{S} \approx 100\GeV$. 

\subsection{$\uude$ direction}

 This will be similar to the $\udd$ direction, except that the 
non-renormalizable terms will 
correspond to the d=4 superpotential term $\frac{\lambda}{M_{p}}\uude$.
 Thus the scalar potential in this case will
have the form 
\be{uddesp} U(\phi) \approx m_{o}^{2} 
\left(1 + K \log\left(\frac{|\phi|^{2}}{M_{X}^{2}}\right) \right) |\phi|^{2} 
+ \frac{\lambda^{2}}{M_{p}^{2}} |\phi|^{6} ~.\ee
The value of $|\phi|$ at the minimum of $U(|\phi|)/|\phi|^{2}$ is given by 
$|\phi|_{o} \approx (|K| m_{S}^{2} M_{Pl}^{2})^{1/4} \approx 3\times
10^{10} |K|^{1/4}\GeV$
for $m_{S} \approx 100\GeV$. 

\section {\bf Q-ball solutions} 
\subsection{Q-ball equation of motion}
  
                From the point of view of cosmology
 and phenomenology, the important quantites 
are the energy and radius of the Q-ball
as a function of its charge $Q$. The
 Q-ball solution is of the form 
\be{q-ball}   \phi = \frac{\phi(r)}{\sqrt{2}} e^{i \omega t}  ~.\ee
 From now on $\phi$ will refer to $\phi(r)$, which is real and positive. 
The energy and charge of the Q-ball are then given by \cite{cole}
\be{energy}  E = \int d^{3}x \left[ \frac{1}{2}
 \left(\frac{\partial \phi}{\partial r}\right)^{2} 
+ U(\phi)  \right] + \frac{1}{2} \omega Q
~\ee
and 
\be{charge}  Q =   \int d^{3}x  \omega \phi^{2}  ~.\ee
The equation of motion for a Q-ball of a fixed value of $\omega$ is given by
\be{eqball}  \phi^{''} 
+ \frac{2}{r} \phi^{'} =  \frac{\partial U(\phi)}{\partial \phi}
- \omega^{2} \phi  ~\ee
where $\phi^{'} = d\phi/dr$. 
We require a solution such that $\phi(0) = \phi^{'}(0) = 0$ and $\phi \rightarrow
0$ as $r \rightarrow \infty$. This corresponds to a tunnelling solution for the potential 
$-\overline{U}(\phi)$ \cite{cole}, where 
\be{ubar} \overline{U}(\phi) = U(\phi) - \frac{\omega^{2}}{2} \phi^{2} ~.\ee 
In practice we vary $\phi(0)$ with these boundary conditions 
until the correct form of solution is obtained 
for a given $\omega$. The energy and charge of the
solution are then calculated using the above expressions. 

\subsection{$H_{u}L$ direction}
         
            The scalar potential which describes the
 L-ball solution for the $H_{u}L$ direction
will be that due to the rapid 
change in the sign of the mass squared term of the effective L-ball field at 
$|\phi| \lae 1 \TeV$. 
We will model this by 
\be{l-pot} U(\phi) \approx \frac{m^{2}}{2}(2 e^{-s \phi} - 1) \phi^{2}    ~,\ee
where $s \approx 1 TeV^{-1}$.
It is straightforward to show (and confirmed by numerical solution of the 
Q-ball equation, \eq{eqball}) that the L-balls in this case will
 correspond to thick-walled L-balls, 
with radius $ r_{o} \approx m^{-1}$ and charge $L \lae  (s m)^{-2}$, 
which, for typical value of $s$ and $m$ will
not be much larger that $10^{3}$. 
This can be understood by noting that 
the effective tunnelling potential in this case is given by 
\be{efbounce} -\overline{U}(\phi) =  \left( \frac{\omega^{2}}{2}
+ \frac{m^{2}}{2} - m^{2} \exp(-s \phi) \right)\phi^{2} ~.\ee   
Thus as $\phi \rightarrow 0$, this tends to
 $ \frac 12({\omega^{2}} - {m^{2}}) \phi^{2} $. A tunnelling solution can only exist
if $w^{2} \leq  m^{2}$. In addition, the largest value of $\phi$ cannot be much
 larger than $ s^{-1}$. From this we can see that
the largest charge of a L-ball of volume V has an upper bound
 $L \lae m s^{-2} V$. From the
 equation of motion it is easy to see that any solution of the L-ball equation will
tend to zero once $r \gae m^{-1}$. Thus $L \lae (sm)^{-2}$ in general.
Therefore the L-balls have a maximum charge
 and an essentially fixed radius in this case, with
 the scale set by the mass scale of the soft SUSY breaking terms. This is in 
contrast with the case of Q-balls which can have a thin-walled solution, for which the 
Q-ball charge is proportional to its volume for sufficiently large charge 
\cite{cole}.

\subsection{$\udd$ and $\uude$ directions}

        In these cases we consider the solution of the Q-ball
 equation for the potentials of \eq{uddsp}
and \eq{uddesp}. 
We find that the B-balls can have a thin-wall solution for sufficiently large charge, 
with a wall thickness of the order of $\omega_{o}^{-1}$. This
 can be understood as follows. 
For the case of the $\udd$ potential, the equation for the B-ball is given by 
\be{B-eq}   \phi^{''} + \frac{2}{r} \phi^{'} = - \omega_{o}^{2} \phi 
+ m_{o}^{2} \phi K \log \left( \frac{\phi^{2}}{\phi_{c}^{2}}\right)
 +  \left(\frac{10 \lambda^{2}}{32}\right)
\frac{\phi^{9}}{M_{p}^{6}}    ~,\ee
where $\omega_{o}$ is defined by 
\be{omo} 
\omega_{o}^{2} = \omega^{2} -  m_{o}^{2} \left[ 1 +
 K \left( 1 + \log \left(\frac{\phi_{c}^{2}}{M_{X}^{2}}\right)\right)\right] 
~\ee
and $\phi_{c}$ is the value of $\phi$ for which the right-hand side of 
\eq{B-eq} vanishes.
 If the initial value of $\phi$ is equal to 
$\phi_{c}$, then $\phi$ will remain constant for all $r$. However, 
if $\phi$ is slightly smaller than 
$\phi_{c}$, then $\phi$ will decrease slowly with increasing $r$ until 
the change 
 $\delta \phi$ is of the order of $\phi_{c}$, after which $\phi$ will 
quickly decrease to
zero as $r$ changes by $\delta r \approx \omega_{o}^{-1}$ (this gives 
the thickness of the wall
 of the B-ball). The value of $r$ at which
$\phi$ begins to rapidly decrease, $r_{c}$, may be estimated by 
perturbing the B-ball equation around $\phi_{c}$. We find that 
\be{twr}  r_{c} \approx \frac{1}{2 \; \omega_{o}}
 \log\left( \frac{\phi_{c}}{\delta \phi(0)}\right)  ~.\ee
Thus, for sufficiently small $\delta \phi(0)$, the B-ball can be made
 as large as we wish. This corresponds to the 
thin-wall Q-ball property that the volume of the Q-ball is proportional
 to its charge. The value of $\omega_{o}$ 
is determined by the solution of the B-ball equation for a given B.
From solving numerically for the B-ball we find that 
$\omega_{o}$ is approximately $4 |K|^{1/2} m_{o}$ for $|K| \lae 0.1$. 
Therefore, so long as the charge of the 
B-ball is sufficiently large that its radius in the thin-wall limit, \eq{twr}, is large
 compared with $\omega_{o}^{-1}$, we can use the thin-wall
expressions. These are given by \cite{cole}
\be{eqtw} \frac{E}{Q} = \left(\frac{2 U(\phi_{o})}{\phi_{o}^{2}}\right)^{1/2}    ~\ee
 and 
\be{qvtw} V =    \frac{Q}{ (2 \phi_{o}^{2} U(\phi_{o}))^{1/2}}   ~,\ee
where $\phi_{o}$ is the value of $\phi$ at the minimum of $U(\phi)/\phi^{2}$.
If the charge of the B-ball is such that 
the radius from equation \eq{twr} is not larger than 
$\omega_{o}^{-1}$, then the B-ball will be thick-walled. In this case, from 
numerical solutions of the B-ball equation, we find that
as B decreases its radius 
stays more or less constant at roughly $4/\omega_{o}$, 
but that the value of $\phi$ inside the B-ball decreases.
This is as we would expect, since the width of a bubble wall is 
purely determined by the mass of the associated scalar and we are effectively
 solving for a bubble here. The energy per unit charge in the thick-walled case 
is found numerically to be larger than in the
 thin-walled case. (In both cases it is not very much 
smaller than the free $\phi$ mass, in contrast 
with the case of gauge-mediated SUSY breaking
where it is proportional to $Q^{-1/4}$ \cite{ks}).
We may estimate the charge $Q_{c}$ at which a Q-ball becomes thin-walled 
by using the thin-wall expression 
for the Q-ball volume as a function of its charge, 
equation \eq{qvtw}, with the radius equal to
 $(|K|^{1/2}m_{o})^{-1}$. For the d=6 $\udd$ potential
\eq{uddsp}, taking $m_{o} \approx 100\GeV$, 
we find that $Q_{c} \approx 10^{26}|K|^{-5/4}$, whilst
 for the d=4 $\uude$ potential, \eq{uddesp}, 
we find that $Q_{c} \approx 10^{17}|K|^{-1}$.  

      The fact that the B-balls in these cases can be thin-walled is important, as it implies
that there is no upper bound on the B-ball charge. This allows for the possibility of 
large charge and so relatively stable B-balls, which may have 
significant consequences for cosmology.

\section{Q-Ball Cosmology}
\subsection{AD baryogenesis and Q-ball formation}

In general, we expect that the
 soft SUSY breaking mass squared term will receive an
order $H^{2}$ correction, coming from 
non-minimal kinetic terms in the supergravity K\"ahler potential \cite{hc}. Such terms 
would be expected in realistic unified theories such as string-type theories. 
The resulting potential for the complex AD field is then of the form 
\be{poth} U(\phi) \approx (m_{S}^{2} - c_{S}H^{2})|\phi|^{2} 
+ \frac{\lambda^{2}|\phi|^{2(d-1)}
}{M_{p}^{2(d-3)}} + ( \frac{A_{\lambda} 
\lambda \phi^{d}}{M_{p}^{d-3}} + h.c.)    ~,\ee
where $A_{\lambda} = A_{\lambda\;o} + a_{\lambda}H$ is the corrected A-term
and $c_{S}$ and $a_{\lambda}$ are typically of the order of 1. In this we are 
choosing the $H^{2}$ correction to have a negative sign in order to have 
 condensate formation and AD baryogenesis.

   The minimal cosmological scenario for AD baryogenesis begins with  
a period of inflation,
during which the Hubble parameter $H$ has a value $H_{I} \approx 10^{14}\GeV$,
 in accordance with COBE perturbations \cite{cobe}.
 The 
inflaton subsequently oscillates coherently about the
 minimum of its potential until 
it completely decays and reheats the Universe to 
a temperature $T_{R}$, which should be less
 than around $10^{9}\GeV$ in order not to thermally 
regenerate too many gravitinos \cite{sarkar}. The
 mass squared term of the AD scalar becomes positive at
 $H \approx m_{S}$, at which time it begins to
 oscillate and the B or L asymmetry is formed 
by the influence of the phase dependent
 A-terms at this time. Throughout this period the 
Universe is matter-dominated by the energy
 of the coherently oscillating inflaton. For completeness, let us note 
that during inflaton oscillation domination 
there will be a thermal background of particles, with temperature $T  \approx 
0.4 (M_{p} H T_{R}^{2})^{1/4}$, coming from inflaton decays \cite{drt}.
 
                  The basic picture for the formation of Q-balls within this scenario 
is as follows. As first observed by Kusenko and Shaposhnikov \cite{ks},
 the AD condensate
can be unstable with respect to growth of space-dependent perturbations of the 
condensate field. This is due to the fact that
when the potential can lead to the formation of Q-balls, it must be increasing 
less rapidly that $\phi^{2}$ for some range of $\phi$. If the condensate starts oscillating
with an initial amplitude lying within this range of $\phi$, then, 
on averaging over the coherent oscillations, 
 it will behave as matter with a negative pressure. This negative pressure
will cause any small perturbations to grow exponentially and go non-linear. 
It is assumed that the further 
collapse of the condensate in the non-linear regime will lead to the formation of
Q-balls, once the non-zero charge (baryon or lepton asymmetry) 
of the condensate comes to stabilize the collapsing
condensate field \cite{ks}. This leaves the
 question of the origin of the initial "seed" perturbations.
Since the AD field should be out of thermal
 equilibrium in order to avoid thermalizing the condensate before
the baryon or lepton asymmetry can be created, the only
 source of initial perturbations in this scenario are 
quantum fluctuations of the AD field created during inflation.   

   Thus, in order to discuss the length scale and charge of the regions of condensate which first
go non-linear and collapse to produce Q-balls, we must first find the spectrum of perturbations 
of the condensate field at $H \approx m_{S}$ due to quantum fluctuations
 and then consider the growth of perturbations of the condensate once $\phi$ begins to
oscillate coherently at $H \approx m_{S}$. 

\subsection{Seed perturbations from quantum fluctuations during inflation}

              Let $\lambda$ be the length scale of a
 perturbation at $H \approx m_{S}$, when the AD
 condensate begins to oscillate coherently. Let
 $\lambda_{e}$ be the corresponding length scale at
the end of inflation. Since the Universe is matter
 dominated after inflation, we have 
\be{01} \frac{\lambda}{\lambda_{e}} =
 \left(\frac{H_{I}}{m_{S}}\right)^{2/3}   ~.\ee
The corresponding length scale at $\Delta N_{e}$ e-folds
 before the end of inflation will be
\be{02} \lambda_{\Delta N_{e}} = \lambda_{e} e^{-\Delta N_{e}}  ~.\ee
This length scale will leave the horizon during
 inflation when $\lambda_{\Delta N_{e}} 
\approx H_{I}^{-1}$. Quantum perturbations of the
 A-D field become classical on leaving the 
horizon, with the magnitude of the perturbation, treating
 $\phi$ as a massless scalar, being given by
\be{03} \delta \phi \approx \frac{H_{I}}{2 \pi}     ~.\ee
Here $\delta \phi$ refers to a perturbation about the minimum 
of the $\phi$ potential with $H > m_{S}$, when $\phi$ has a negative 
mass squared term of the order of $H^{2}$. Thus the physical $\phi$ scalar 
will have a mass of the
 order of $H$. Although this is not really an effectively massless scalar, 
which would require a mass small
 compared with $H$, it is not very massive relative to $H$ and so we expect 
the massless result to be 
roughly correct. Once the perturbation is larger than the horizon it will 
appear to be a
 homogeneous perturbation on sub-horizon scales and so will coherently 
oscillate about the minimum
 of its potential and red-shift like matter, $\delta \phi \propto a^{-3/2}$. 
This assumes 
that the mass is large compared with $H$, so that we can ignore the 
$3 H \dot{\phi}$ 
damping term in the $\phi$ equation of motion. Again this is not strongly 
satisfied, but we expect that
it will roughly give the correct behaviour.
 Thus at the end of inflation the spectrum of 
perturbations will be given by $\delta \phi_{e}
 \approx e^{-3 \Delta N_{e}(\lambda)/2} H_{I}/(2
 \pi)$, where $\Delta N_{e}(\lambda)$ refers to the
 number of e-folds before the end of inflation
 at which the perturbation of length scale $\lambda$
 first leaves the horizon. Suppose the perturbation
 re-enters the horizon at $a_{\lambda}$. Once back inside the
horizon, the perturbation will be space-dependent.
 Its subsequent evolution will depend on whether
its energy density is dominated by the gradient or
 potential energy. The $\delta \phi$ equation of
 motion is
\be{04} \delta \ddot{\phi} + 3 H \delta \dot{\phi}
 - \nabla^{2} \delta \phi = -V^{'}(\delta 
\phi)    ~,\ee
where $V(\delta \phi) \approx H^{2}
 \delta \phi^{2}$. The gradient energy will dominate 
if $ \vec{k}^{2} \gae H^{2}$, where $\delta \phi \propto e^{i \vec{k}
\cdot\vec{x}}$.
 This will obviously be satisfied once the perturbation 
enters the horizon. Since, as the Universe
 expands, $\vec{k}^{2}$ is proportional to 
$a^{-2}$ while $H^{2}$ is proportional to $a^{-3}$, in general the
perturbations will be gradient energy dominated once they 
re-enter the horizon. Therefore
$\delta \phi \propto a^{-1}$. Putting all this together, we find that the spectrum of 
the sub-horizon sized perturbations at $H \approx m_{S}$
due to quantum fluctuations during inflation 
is given by 
\be{spec} \delta \phi(\lambda) \approx
 \frac{1}{2 \pi m_{S} H_{I}^{1/2} \lambda^{5/2}}    ~.\ee

\subsection{Negative pressure and Q-Ball formation}

      The growth of these perturbations once the AD field begins to oscillate coherently can
 be understood by considering the case of B-balls and 
the squark potential for values of $\phi$ somewhat smaller than 
the value of $\phi$ at the minimum of the potential (this describes the potential of the
 oscillating field once $H$ is less than $m_{S}$), 
\be{logm} U(\phi) \approx 
\frac{m_{o}^{2}}{2} \left( 1 + K \log\left(\frac{\phi^{2}}{M_{X}^{2}}\right)
\right) \phi^{2}    ~,\ee
where we typically expect $|K| \approx 0.01-0.1$.
For $|K|$ small compared with 1, this has the form $U(\phi)
 \propto \phi^{2 + 2 K}$. As discussed
in reference \cite{turner}, the equation of state of matter corresponding to a coherently 
oscillating scalar in a potential of the form $\phi^{n}$ is given by
 \be{eqst} p = (\gamma - 1) \rho \;\;\; ; \;\;\;\;\;\; \gamma = 
\frac{2n}{n+2} ~.
\ee
 Thus for a scalar oscillating in the potential given by \eq{logm} 
the equation of state is given by \cite{mcd} 
\be{eqs} p = \frac{K}{2} \rho   ~,\ee
which gives a negative pressure if $K < 0$. Perturbations in the field will 
grow according to 
\be{pertrho} \ddot{\delta}_{\vec{k}} = -\frac{K \vec{k}^{2}}{2} 
\delta_{\vec{k}}
      ~,\ee 
where the perturbation of the density of the AD field on the length scale 
$\lambda = 2 \pi/|\vec{k}|$ is proportional to $\delta_{\vec{k}} = \delta 
\rho_{\vec{k}}/\rho$.
With $\delta \rho \propto 2 \phi \delta \phi$, this implies that 
\be{pert} \delta \ddot{\phi}_{k} = -\frac{K \vec{k}^{2}}{2} \delta \phi  
    ~.\ee 
Thus the quantum fluctuations of the $\phi$ field on the scale $|\vec{k}| 
= 2\pi / \lambda$ will
grow exponentially with time according to
\be{phiexp} \delta \phi_{k} = \delta \phi_{i\;k}\; 
\exp\left( \left(\frac{|K| \vec{k}^{2}}{2} \right)^{1/2}t \right) ~,\ee
where $t=0$ corresponds to the beginning of the coherent oscillations and 
$\phi_{i}$ and $\delta \phi_{i}$ are the 
values of the field and its perturbation respectively at $H \approx m_{S}$. 
In this we have neglected the expansion of the Universe for simplicity. 
This is justified as we 
need only establish the value of $H$ at which the perturbations go non-linear 
in an expansion time $H^{-1}$. Since $\delta \phi_{i}$ and $\phi_{i}$ 
scale the same way due the expansion of the Universe, we can calculate 
their ratio at $H \approx m_{S}$.

        How can B-balls form as a result of this negative pressure? 
The naive picture is that 
a perturbation on some scale $\lambda$ will go non-linear once 
$t \gae m_{S}^{-1}$ (which is the smallest time scale on which
 B-balls can form and on which we can
 average over coherent oscillations to obtain a negative pressure), 
causing the AD condensate to collapse
 into fragments of size $\lambda$, trapping 
some charge B, which will then form into
 B-balls with a charge of the order of B (depending on the
efficiency of B-ball formation in the non-linearly
 collapsing fragments). However, by considering
 only
the effect of negative pressure we are neglecting
 the effect of the charge B. The charge of the 
B-ball serves to prevent the soliton from 
collapsing to a radius smaller than the B-ball radius. 
Therefore we do not expect perturbations to be able
 to grow on length scales smaller than the 
B-ball radius once charge is taken into account.
 We can use the negative pressure 
arguement once the length scale going non-linear
 at a time $t$ is larger than the final B-ball 
radius. We would then expect the B-balls to
 form quite efficiently when these B-ball size perturbations
go non-linear. The charge of the B-ball will be given by the charge contained in 
a volume of radius of the order of the B-ball radius at this time. 

                 The time at which a perturbation of scale
 $\lambda$ goes non-linear is given by
\be{tnl} t \approx 
\frac{\alpha_{k}}{2 \pi} \left(\frac{2}{|K|}\right)^{1/2} \lambda    ~,\ee
where 
\be{alpha} \alpha_{k} = \log\left(\frac{\phi_{i}}{\delta \phi_{i\;k}}\right)   ~.\ee
We find that $\alpha_{k} \approx 34 \; (44)$ for the d=4 (d=6) directions. 
In practice, the B-balls will typically
turn out to be thick-walled, with radius $r_{o} \approx
 (|K|^{1/2} m_{S})^{-1}$. Perturbations on this scale,
 which have the largest possible growth in a time 
$H^{-1}$,
 will go non-linear at
\be{tq}  t \approx \frac{10}{|K| m_{S}}   ~,\ee
corresponding to $H \approx 0.1 |K| m_{S}$.
 In order to find the charge of the 
resulting B-balls we must find the charge density of the Universe at this time. We will 
assume that the charge asymmetry
 corresponds to the presently observed baryon asymmetry, 
$\eta_{B} \approx 10^{-10}$.
Then the charge asymmetry at a given value
 of $H$ during the inflaton oscillation dominated 
era prior to reheating is given by
\be{qasy}    n_{B} \approx \left( \frac{\eta_{B}}{2 \pi}\right)  \frac{H^{2}
 M_{p}^{2}}{T_{R}} ~.\ee
Thus we find that the charge of 
the region with radius $r_{o} \approx (|K|^{1/2}m_{S})^{-1}$ is given by
\be{chasy} B \approx 10^{15} |K|^{1/2} 
\left(\frac{\eta_{B}}{10^{-10}}\right)
\left(\frac{10^{9}\GeV}{T_{R}}\right)
\left(\frac{100\GeV}{m_{S}}\right)
~.\ee
Thus, with $|K| \gae 0.01$, we see that
B-balls of charge larger 
than $10^{14}$ are likely to be formed, depending
 on the reheating temperature and the efficiency
 with which this charge is trapped within the B-balls. 

            For the case of L-balls, we have seen that
 L-balls have a maximum charge of the order of 
$10^{3}$ and field strength of the order of $1\TeV$, which
 is much smaller than the initial amplitude of the d=4 AD field at 
$H \approx m_{S}$. This suggests that L-balls
 cannot be formed by the collapse of an
unstable condensate at $H \approx m_{S}$ and that AD baryogenesis
 along the $H_{u}L$ direction
 will be essentially unaltered from the conventional scenario. 

\subsection{Cosmological consequences of Q-balls in the MSSM}

           Given that B-balls with a large charge will naturally form
 in the AD scenario along the squark directions,  
what might be their consequences for cosmology? This is crucially
 dependent upon the temperature at which the 
B-balls decay. If they decay after the electroweak phase transition
 has occured at $T_{ew}$, then we will have the possibility of 
a new version of AD baryogenesis, in which the baryon asymmetry
 originates from the decay of the B-balls rather than the 
decay of the condensate. For example, if we were to consider L violating
 interactions which are in thermal equilibrium
together with anomalous B+L violation at $T \gae T_{ew}$
 (as may be expected in extensions of the MSSM with 
light Majorana neutrinos via the see-saw mechanism), then any B asymmetry
 coming from thermalization or decay of the AD condensate
would be subsequently washed out \cite{ellis}.
However, the B contained in the B balls will not be erased, at least for
 some range of reheating 
temperatures, due to the
fact that 
the large field inside the B-ball prevents thermal particles from
 penetrating beyond the surface of the B-ball, thus
suppressing the thermalization rate \cite{ks}. To see this, 
 we will give a conservative 
upper bound on the reheating temperature from requiring that
 the B-balls are not thermalized.
Thermal particles will penetrate into the B-ball down to a stopping radius 
$r_{st}$, corresponding to 
$g \phi_{st} \approx T$, where $\phi_{st} = \phi(r_{st})$ and where the thermal
 particles gain a mass
 $g \phi$ from interacting with the B-ball. Let us also assume that the
 thermal particles reflect from
the "hard" region of the B-ball corresponding to $r < r_{st}$. Then there
 will be a flux of thermal 
particles flowing in and out of the "soft" part of the
 B-ball at $r > r_{st}$. The most effective 
way possible for this flux to remove charge from the B-ball would be via
 B absorbing inverse decay 
processes with a $maximum$ possible rate $\Gamma_{inv\;d} = k_{d} T$, 
where $k_{d} \approx 10^{-2}$ for strong
interactions 
\cite{enqotto}. The true inverse decay 
rate can be significantly smaller, depending on the details of the model.  
For large enough $r$, the thick-walled B-ball  will be
   roughly described by a Gaussian profile, $\phi(r) \approx \phi_{o} e^{-r^{2}/r_{o}^{2}}$.
 Therefore the stopping radius will be given by 
$r_{st} \approx 
r_{o} log^{1/2} (\phi_{o}/\phi_{st})$. 
Thus, since the charge density 
at radius $r$ is $\omega \phi^{2}(r)$, which
 will be approximately constant over a change in radius 
$\delta r \approx r_{o}^{2}/(2 r_{st})$, we find that the $maximum$
 rate of loss of charge by the B-ball via inverse decays
 is approximately given by 
\be{qbtherm} \frac{1}{B} \frac{dB}{dt}
 \approx - \frac{3}{2} \frac{\phi_{st}^{2}}{\phi_{o}^{2}}
\frac{r_{st}}{r_{o}} k_{d}T
  ~,\ee
where $r_{o} \approx 
 (|K|^{1/2}m_{S})^{-1}$ is the radius within which most of the
 Q-ball charge is concentrated.
The condition for the Q-ball to survive
 thermalization is then that $\dot{B}/B \lae H$. 
This gives an upper bound on the reheating temperature
\be{thermb}   T_{R} \lae \frac{2 g^{2} k_{T} \phi_{o}^{2}}{3 k_{d}M_{p} log^{1/2}
\left(\frac{g\phi_{o}}{T}\right)}   ~.\ee
For thick-walled B-balls, the value of the scalar field at the centre of the B-ball, 
$\phi_{o}$, is given by $\phi_{o} \approx 0.5 |K|^{3/4} m B^{1/2}$, where 
$B$ is given by \eq{chasy}. Thus, with
$g \approx 1$ and $k_{d} \approx 10^{-2}$ for strong interactions and
with $m \approx 100GeV$, we obtain an upper bound $T_{R} \lae 
5 \times 10^{5}|K|GeV$. 
Thus, although a
 non-trivial upper bound 
may exist, even with relatively conservative assumptions there is still a range
 of reheating temperatures 
for which the B-balls can survive thermalization. Clearly a much more detailed analysis of 
thermalization is required to give the true upper bound on $T_{R}$, which could
 easily be weaker than our conservative estimate. (We have checked that destruction of the 
B-ball by direct collisions of thermal particles with the
 hard region is less important than thermalization 
of the soft region \cite{qbnew}). 

               From this we may conclude that, at least for some range of $T_{R}$, a B asymmetry
can be preserved in the presence of rapid L violating interactions 
so long as the B-balls decay at $T_{d} < T_{ew}$. 
Another interesting possibility is connected with the 
d=4 B-L conserving directions $\uude$ and $QQQL$. These cannot
 produce a B asymmetry in conventional AD baryogenesis, 
since the condensate will be thermalized at temperatures large compared with
 $T_{ew}$ and so anomalous B+L violation 
will wash out the B asymmetry. However, if the Q-balls, which in this case are charged
 under $U(1)_{B+L}$, decay after 
the electroweak phase transition, then a non-zero B asymmetry will
 result, together with an equal L asymmetry. 
In addition, even in those cases where conventional AD baryogenesis can, in
 principle, account for the observed baryon
asymmetry, the B-ball decay mechanism might replace the conventional 
mechanism if, after the collapse of the AD condensate,
 most of the asymmetry were trapped in the B-balls.

                    The Q-ball decay rate to light fermions is proportional to the area of the Q-ball. 
It has been estimated to have an upper bound, which is likely to be saturated for Q-balls 
with $\phi_{o}$ much larger than $m_{o}$, given by \cite{cole2} 
\be{decay}    \frac{dQ}{dt} = -\frac{\omega^{3}A}{192 \pi^{2}}   ~,\ee
where $A$ is the area of the Q-ball and 
where $\omega \approx m_{o}$ for $|K|$ small compared with 1.
 The decay rate will depend on 
whether the Q-ball is thin or thick-walled. For thin-walled Q-balls 
the area of the Q-ball is related to the charge by
\be{qatw} A = \frac{(36 \pi)^{1/3}Q^{2/3}}{k^{2/3}}    ~,\ee
where $k = (2 \phi_{o}^{2} U(\phi_{o}))^{1/2}$.
Thus the lifetime of the Q-ball in this case is given by
\be{lttw} \tau = 144 \pi \left(\frac{4 \pi}{3}\right)^{2/3} 
\frac{k^{2/3}Q^{1/3}}{\omega^{3}} ~.\ee
For the thick-walled case, the area of the Q-ball is independent of 
its charge, being fixed by its radius $r_{o} \approx (|K|^{1/2}m_{o})^{-1}$. 
The Q-ball lifetime in this case is then given by
\be{ltfw} \tau = \frac{48 \pi Q}{r_{o}^{2} \omega^{3}}   ~\ee
For charges less than of the order of $10^{26}|K|^{-5/4}$ for the 
$\udd$ direction and $10^{17}|K|^{-1}$ for the 
$\uude$ direction, the Q-balls will be thick-walled. The temperature at 
which the Q-balls decay is given by, assuming radiation domination,
\be{tempd} T_{d} = \left(\frac{1}{k_{T}}\right)^{1/2} 
\left(\frac{M_{p}}{2 \tau}\right)^{1/2}
~,\ee
where $k_{T} = \left(\frac{4 \pi^{3} g(T)}{45}\right)^{1/2}$.
Thus, assuming $\omega \approx m_{o} \approx 100
\GeV$,  thick-walled Q-balls will decay at 
\be{ttw}  T_{d} \approx \frac{15}{|K|^{1/2}}  
                   \left(\frac{\omega}{100 \GeV}\right)^{1/2}
                   \left(\frac{10^{15}}{Q}\right)^{1/2} \GeV 
~,\ee
using $k_{T} \approx 17$. Thus we see that the Q-balls will decay 
at a temperature 
less than 100 GeV if $Q \gae 2 \times 10^{13}|K|^{-1}$. For thin-walled Q-balls, 
the right-hand side of equation \eq{ttw} has an additional factor
 $\left({Q}/{Q_{c}}\right)^{1/3}$, where $Q_{c}$ is the value of $Q$ at 
which the thin-wall limit becomes valid. Q-ball decay is  
summarized in Fig. 2. 
\begin{figure}
\leavevmode
\centering
\vspace*{80mm} 
\includegraphics{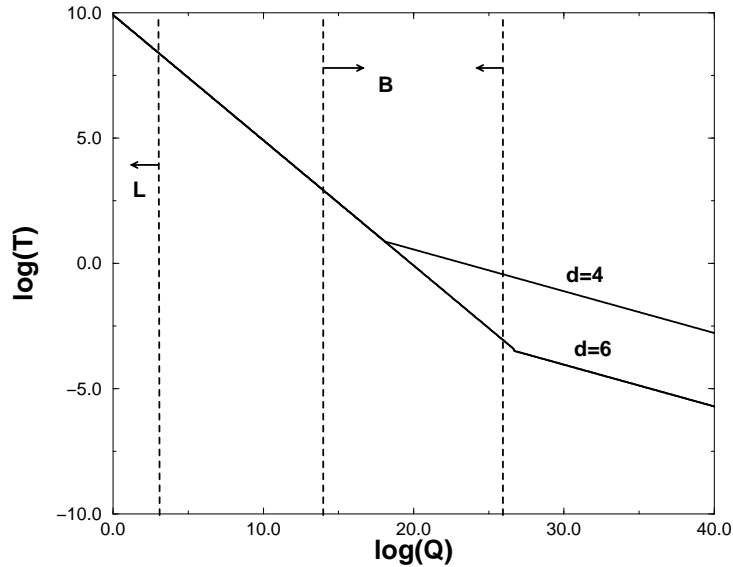}   
\caption{Q-ball decay temperature $T$ vs. the charge $Q$. The regions
where L-balls and B-balls exist are also indicated.}
\label{kuva2}       
\end{figure} 
Since we have shown that it is likely that the B-balls will have a charge greater than 
$10^{14}$ in the MSSM, we see that it is
 quite natural for MSSM B-balls to decay after the electroweak 
phase transition has occured. Thus it is possible to generate the observed B asymmetry
 via the decays of B-balls occuring after the
 electroweak phase transition. This is true even in those
 cases with rapid L violating interactions or B-L conservation, where the conventional AD
 mechanism fails.

                  The decay temperature depends on $B$, which in
 turn depends on the reheating temperature $T_{R}$. 
Nucleosynthesis requires that $T_{R} \gae 1\MeV$, which
 imposes an upper limit on the possible charge of the B
balls, $B \lae 10^{26}$. This in turn imposes a lower bound on 
$T_{d}$. For the d=4 case we find that $T_{d} \gae 
50\MeV$, whilst for the d=6 case we find $T_{d} \gae 50\keV$ (see Fig. 2).  
 In principle, B-balls could decay at temperatures as low as
 that of the quark-hadron phase transition or less. This could have interesting
 consequences for the 
spatial distribution of baryons and inhomogeneous nucleosynthesis \cite{inhom}. 

            L-balls, which have $Q \approx 1000$,
would decay at $T_{d} \approx 10^{7}\GeV$
 even if they could be formed primordially.
 It is therefore unlikely that such L-balls 
could have any cosmological consequences.
 However, it is possible that L-balls, which
 have a field strength of the order of 1 TeV or less, 
could play a role in the physics of the electroweak phase transition.  

\section{Conclusions}

                  We have shown that it is quite natural for Q-balls to exist in the MSSM without
requiring any special choice of SUSY breaking mechanism. For the
 case of B-balls associated with D-flat 
directions involving squark fields, these are usually sufficiently stable 
to decay after the electroweak phase transition has occured. As a result, the
 Affleck-Dine mechanism 
along these direction will 
have to be revised, as the baryon asymmetry will
come from condensate collapse and B-ball decay 
rather than simply from the decay of a coherently oscillating scalar condensate. 
As many extensions of the MSSM introduce additional L violating interactions, which, 
if they are sufficiently 
rapid to be in thermal equilibrium when anomalous B+L violation is in 
thermal equilibrium, rule out 
conventional AD baryogenesis but not B-ball decay baryogenesis,
 the potential importance of B-ball decay baryogenesis in the MSSM and its extensions 
is clear. 
We have also pointed out the existence of L-balls of small charge, characterized by a field 
strength not much larger than a TeV at most. Although such L-balls
 will be very short-lived, decaying at a
temperature of around $10^{7}\GeV$ even if produced primordially, and so
 are unlikely to 
significantly alter the conventional Affleck-Dine mechanism, they
might nevertheless be produced during the electroweak phase transition.

             There are several points that this discussion of Q-balls in the MSSM has not 
addressed. The details of the formation of B-balls from the non-linear collapse of the
 condensate
and the efficiency which which charge is trapped in B-balls needs to be clarified, in order to 
obtain an accurate estimate of the B-ball charge, which determines the temperature at
 which the B-balls decay. 
The physics of the thermalization of B-balls should also should be analysed in detail, in
 order to obtain constraints 
on the parameters of the models, in particular the reheating temperature. 
Given that the B-balls are decaying after the electroweak phase transition has
 occured, we should also 
consider what are the possible consequences of B-ball decay for other issues in cosmology.
If the B-ball has a sufficiently large charge, corresponding to a
 low enough reheating temperature after inflation, 
then it
 might decay at a temperature close to that of the quark-hadron phase transition or 
nucleosynthesis. This could have consequences for inhomogeneous nucleosynthesis models. 
B-ball decay might also have consequences for dark matter in SUSY models,
 given that the B-balls are made of squarks which will produce neutralinos when they decay.
 For the case of L-balls, it would be interesting to consider the 
possible consequences of L-ball production during the electroweak phase transition.
 These issues will be the subject of future work.
 
\subsection*{Acknowledgements}   This work has been supported by the
 Academy of Finland and by a Marie Curie Fellowship under EU contract number 
ERBFM-BICT960567.

\end{document}